# Percolative superconductivity in La$_2$CuO$_{4.06}$ by lattice granularity patterns with scanning micro X-ray absorption near edge structure

Nicola Poccia[1,2], Matthieu Chorro[3], Alessandro Ricci[4,2], Wei Xu[5], Augusto Marcelli[6,7,2], Gaetano Campi[8,2], Antonio Bianconi[2,8]

[1] MESA+ Institute for Nanotechnology, University of Twente, P. O. Box 217, 7500AE Enschede, Netherlands.

[2] RICMASS Rome International Center for Materials Science Superstripes, via dei Sabelli 119A, 00185 Roma, Italy.

[3] Synchrotron SOLEIL L'Orme des Merisiers, 91190 Paris S.Aubin, France.

[4] Deutsches Elektronen-Synchrotron DESY, Notkestraße 85, D-22607 Hamburg, Germany

[5] Beijing Synchrotron Radiation Facility, Institute of High Energy Physics, Chinese Academy of Sciences, 100049, P. R. China

[6] Istituto Nazionale di Fisica Nucleare - Laboratori Nazionali di Frascati, 00044 Frascati, Rome, Italy

[7] NSRL, University of Science and Technology of China, Hefei, 230026, P.R. China

[8] Institute of Crystallography, CNR, via Salaria Km 29.300, Monterotondo, 00015 Rome, Italy.

**Abstract**

The simplest cuprate superconductor La$_2$CuO$_{4+y}$ with mobile oxygen interstitials exhibits a clear phase separation, but only recently a bulk multiscale structural phase separation has been observed by using scanning micro X-ray diffraction. Here we get further information on the structural phase separation, using local probe X-ray absorption near edge structure. The spatial distribution of superconducting units is a key parameter controlling percolative superconductivity in complex matter with dispersed superconducting units. These oxides form super-molecular architectures made of superconducting atomic monolayers intercalated by spacers. Oxygen interstitials enter into the rocksalt La$_2$O$_{2+y}$ spacer layers forming oxygen interstitials rich puddles and poor puddles. Their spatial distribution has been determined by using scanning La L$_3$-edge micro X-ray absorption near edge structure. Percolating networks of oxygen rich puddles are observed in different micrometer size portions of the crystals. Moreover, the complex surface resistivity shows two jumps associated to the onset of intra-puddle and inter-puddles percolative superconductivity. The similarity of oxygen doped La$_2$CuO$_{4+y}$, with the well established phase separation in iron selenide superconductors is also discussed.

**Keywords:** defects, oxygen interstitials, disorder, puddles, phase separation superstripes, inhomogeneous superconductors, cuprates, X-ray absorption, XANES.

Complex lattice architectures composed of transition metal oxides display a rich variety of coexisting nanoscale electronic and magnetic phases going from superconductor, multiferroic, and colossal magnetoresistance units [1,2]. In this soft electronic matter, a new low energy physics that goes beyond the standard condensed matter is emerging opening new perspectives to technology and nanotechnology [3]. The control of the spatial organization of defects in transition metal oxides is a key issue in this field [5,6]. The study of the multiscale phase separation from nanoscale to micron scale, demands the development of new bulk experimental methods. Indeed, standard experimental probes, like standard scanning tunneling microscopy, fail to unveil the existing competition among nano-scale phases since provides information only on the sample surface and a few layers behind [7]. On the contrary, standard experimental techniques looking at the k-space like X-ray diffraction, neutron diffraction and angular resolved photoemission provide non-local and space averaged information. Actually, local structure experimental probes, like Pair

Distribution Function (PDF) analysis of neutron diffraction [8,9] or X-ray Absorption Near Edge structure (XANES) [10-15] allow to study nanoscale structures, without time or spatial averaging. These latter local methods, applied to complex high-$T_c$ superconductors [16-22], have provided evidence of local electronic and lattice fluctuations otherwise not accessible. Furthermore, local lattice fluctuations in the $CuO_2$ plane and the electronic phase separation have been shown to favor high temperature superconductivity (HTS) in cuprates [23-36].

In the proposed superstripes scenario [25-29], the local lattice fluctuations at atomic scale [16-24] form striped puddles at nanoscale, and these puddles form complex inhomogeneous textures [25-29], observed up to the micron scale [5,6]. Therefore the superstripes scenario is a particular case of percolative superconductivity [37-54], where superconductivity emerges in a network of striped superconducting units in a two-dimensional atomic layer. In this scenario the spatial distribution of superconducting units is a key issue.

Percolation of superconducting puddles has been recently observed also in iron selenide superconductors [55-58]. The scientific agreement on the structural phase separation in $R_yFe_{2-x}Se_2$ (with R=K,Rb,Cs) has been reached rapidly, using experimental methods not available 20 years ago. Here different magnetic domains and networks of metallic domain walls are now recognized as an essential feature for HTS.

Electronic phase separation in strongly correlated superconductors is predicted by multiband Hubbard models near a Lifshitz transition [59] and confirmed by time dependent phenomena [60,61].

Advances in the control of the nanoscale phase separation by tuning lattice modulations and defects multiscale organization are now of a high interest in strongly correlated materials. Recently the distribution of lattice defects in HTS has been found to play a key role in the superconducting phase [62-68], in fact the distribution of defects controls the electronic features in the range of 10-100 meV near the Fermi level and therefore multi-gaps superconductivity [69-74].

$La_2CuO_{4+y}$ is a prototype of cuprate superconductors, showing a clear phase separation. Early neutron powder diffraction studies have shown the occurrence of a phase separation for y<0.055 [75] and the average location of the oxygen interstitials (O$i$) [76]. A further neutron powder diffraction work [77] has established a phase diagram for $La_2CuO_{4+y}$ with 0<y<0.055, demonstrating a miscibility gap between an antiferromagnetic phase at y~0.01 and a superconducting phase at y~0.055. The phase separation for y>0.055 was object of discussion for long time. Neutron diffraction on single crystals at higher concentration of O$i$ [78-80] have provided evidence for superstructure satellites, but still there is a lack of information on the spatial location of multiple phases. An extended X-ray absorption fine structure experiment on a y=0.1 sample has shown local lattice fluctuations in the $CuO_2$ plane [81]. Moreover, the O$i$ have a large mobility in this system since the spacer layers have a large tensile strain due to lattice misfit between the active and spacer layers [82-85]. The spatial distribution of puddles with 3D ordered O$i$ has been described by scanning micro X-ray diffraction in an optimally doped $La_2CuO_{4.1}$ ($T_c$ = 40 K) [62] and in an underdoped $La_2CuO_{4.06}$ [63,64]. These experiments pointed out that O$i$-rich puddles (ORP) form a scale-free network of superconducting puddles at the optimum doping, which favors HTS.

Here we have investigated the statistical distribution of O$i$ in the underdoped $La_2CuO_{4+y}$ using scanning X-ray Absorption Near Edge structure (μXANES) [86]. This approach is complementary to scanning micro X-ray diffraction (μXRD) previously applied to investigate multiscale structural phase separation in these materials [62-66].

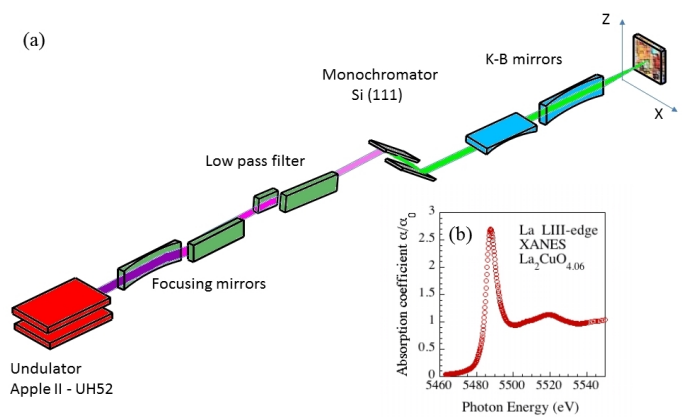

**Figure 1.** *a)* Experimental setup for μXANES measurements, performed on the dedicated-beamline LUCIA at SOLEIL. The monochromatic X-ray beam is focused on a selected spot of 2.5x2.5 μm$^2$, *b)* A typical La $L_3$-edge XANES spectrum recorded in fluorescence mode on a single micrometer size spot.

The advantage of using XANES, instead than X-ray diffraction, consists in the possibility to probe the local lattice structure within a nanoscale cluster centered at the selected absorption site. Indeed µXRD is a non-local probe of the k-space and it is sensitive only to puddles of ordered O$i$ over a scale larger than the diffraction coherence length. Therefore, µXANES is a unique powerful tool for the investigation of complex multiphase materials characterized by either local and spatial sensitivity with no space averaging and with a femtosecond time response [12-15]. We report µXANES experiments providing accurate information on the local atomic structure of an underdoped $La_2CuO_{4.06}$ sample. XANES spectra have been collected at the Lanthanum $L_3$-edge [25-29], to probe directly the local structure within ~0.8 nm around the selected La atomic species in the $La_2CuO_{4+y}$ spacer layer.

We have examined a $La_2CuO_{4+y}$ single crystal, with the space group F$mmm$, in the underdoped regime with y = 0.06, which corresponds to an electronic doping of 0.1 holes per Cu site. The sample was grown first as $La_2CuO_4$ by the flux method and then it was doped by electrochemical oxidation. Details on sample preparation and characterization have been reported elsewhere [63].

The µXANES measurements were performed on the beamline LUCIA at SOLEIL [70] in Paris. The scheme of the experiment is shown in Fig. 1. The X-ray source is an electron undulator Apple II-UH52, operating in the range of soft X-rays (0.8-8 keV) with high brilliance (~1.6*10$^{11}$ph/s/400 mA) equipped with a double Si-111 monochromator. The monochromatic photon beam is focused on a spot of 2.5x2.5 µm$^2$ with a Kirkpatrick-Baez mirrors apparatus. The sample is mounted on an x-z translator that allows to scan the sample at the nanometer resolution, as shown in panel a) of Fig. 1. At each x-z position, the Lanthanum $L_3$-edge XANES signal was recorded in the fluorescence mode with a 4 elements silicon drift Bruker detector. A typical spectrum corresponding to the electronic transition from the La $L_3$ core level at the binding energy of 5383 eV, is shown in panel b) of Figure 1.

The spectrum is very simple and shows two multiple scattering resonances [10-15] separated by a minimum. These two resonances are determined by the local structure of a nanoscale cluster of neighboring atoms around the La photoabsorber.

Using µXANES, we have collected the space-resolved XANES response, probing the local structure around the La-ion in the spacer layers. This is obtained by recording polarized **E**//**ab** La $L_3$-edge XANES spectra at the $La_2CuO_{4.06}$ crystal at each 2.5x2.5 µm$^2$ illuminated spot. The La $L_3$-XANES spectra were recorded with energy steps of 0.2 eV in the near edge region at 4 s integration (see Fig. 2). We have found that the XANES spectra, collected in different spots are different. Figure 2 (upper panel) shows two typical La $L_3$-edge XANES spectra, collected in two different locations of the sample.

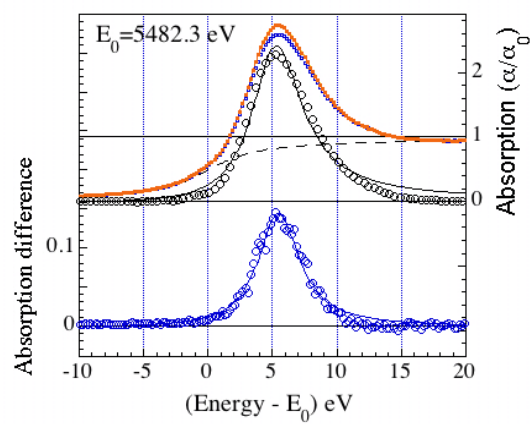

**Figure 2. Upper panel** Two typical La $L_3$-edge XANES spectra (blue and red dots) recorded at two different illuminated spots of the crystal. The black dashed line is the atomic absorption. The difference between the experimental absorption and the atomic absorption is shown by empty black circles (upper panel). This curve is the measure of the partial absorption cross section for final states where the photoelectron is trapped in a nanoscale cluster for a finite time in a shape resonance [10-15]. The black continuous curve is the fit of the shape resonance by a Fano lineshape curve. **Lower panel.** The difference between XANES spectra recorded in two different spots (i.e., at two 2.5x2.5 µm$^2$ pixels) of the sample. The experimental difference is well beyond the error bar and it is like the shape resonance. XANES differences from point to point in the real space

(lower panel) are determined by the variation of the shape resonance from point to point.

The typical spatial variation of the La $L_3$-edge XANES spectra is shown by the XANES difference spectrum in the lower part of Fig. 2, where only a broad peak centered at 5 eV above the ionization threshold appears.

The XANES spectra of condensed matter are determined by core excitons below the atomic ionization threshold and shape resonances above the absorption threshold [10-15]. The typical **E//ab** polarized La $L_3$-edge XANES spectrum shown in Fig. 2 is very simple since no bound excitons are appear below the absorption threshold. Above threshold the absorption is given by

$$\alpha(\omega) = \alpha_a(\omega)[1 + \chi(\omega)] = \alpha_a(\omega) + \alpha_a(\omega)\chi(\omega)$$

where $\alpha_0(\omega)$ is the atomic absorption, due to electronic transitions from the atomic La $(2p)_{3/2}$ core level to the continuum, and $\chi(\omega)$ is the modulation of the atomic absorption cross section $\alpha_0(\omega)$, due to photoelectron quasi-bound final states degenerate with the continuum. Therefore the local lattice structure controls the photoelectron final states probed by $\chi(\omega)\alpha_0(\omega)$.

Figure 2 shows the behavior of $\chi(\omega)\alpha_0(\omega)$ contribution, extracted by subtracting the atomic absorption from the experimental spectrum. The atomic absorption spectrum has been obtained by fitting the experimental XANES spectrum with an arctangent in a range of 10 eV below the absorption edge and in the range between 15 and 20 eV above the absorption edge as it is shown in Fig. 2. The inflection point at $E_0$ of the arctangent line has been selected as the zero of the energy scale. $E_0$ is the ionization potential of the La $(2p)_{3/2}$ core level and therefore is the energy threshold for electronic transition to the continuum. The measured absorption coefficient $\alpha$ has been divided by the value of the atomic absorption spectrum at high energy $\alpha_0$. Therefore $\alpha/\alpha_0$ is the absorption coefficient measured in units of the atomic X-ray absorption jump at the La $L_3$-edge. The extracted modulation factor:

$$\chi(\omega)\alpha_0(\omega) = \alpha(\omega) - \alpha_0(\omega)$$

has structural information and is shown in the upper panel of Fig. 2. This spectrum shows a single broad asymmetric peak centered at energy 5 eV. This broad peak is clearly interpretable as a shape resonance [11,12], which is due to a quasi-bound final state where the excited photoelectron is confined in a nanoscale cluster [10-15]. This quasi-bound state arises by a multiple scattering resonance (MSR). The configuration interaction between the MSR quasi-bound state and the continuum is expected to give a spectral Fano line-shape. Therefore the Fano line-shape is a fundamental feature of MSR in XANES spectra [10-15]. This interpretation is confirmed by the good fit with a Fano line-shape of the MSR, shown in the upper part of Fig. 2.

The lower panel in Fig. 2 shows that the difference of the two XANES spectra recorded in different points shows a similar Fano line-shape as the shape resonance in the upper panel. It is well established that $\alpha_0$ being the atomic cross section should not show any spatial dependence [10-15], therefore the difference between the two XANES spectra shown in the lower panel of Fig. 2 is due to $\alpha_2(\omega) - \alpha_1(\omega) = \alpha_a(\omega)\chi_1(\omega) - \alpha_a(\omega)\chi_2(\omega)$, which is determined by the difference of the MSR, changing from spot to spot.

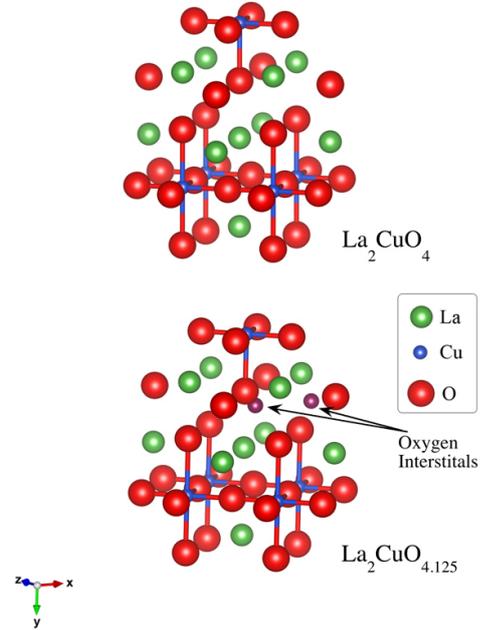

**Figure 3.** The cluster of neighboring atoms surrounding the La absorbing atom in a $La_2CuO_{4+y}$ crystal that determines the shape resonance in the La $L_3$-XANES. The photoelectron wave emitted by the absorbing atoms, is scattered by neighboring atoms Cu (blue spheres), oxygen (red spheres) and La (green spheres) ions. The size of the cluster is 8 Å. In oxygen rich puddles (like in the local structure of the $La_2CuO_{4.125}$ crystal) the cluster includes three additional O*i* (indicated by the black arrows).

The raw data of the difference XANES spectra clearly indicate a relevant spatial variation of the local structure: the cluster of neighbor atoms surrounding the absorbing La-ion.

The determination of the nanoscale cluster size, probed by the excited photoelectron in the XANES spectra, is the first step for its quantitative interpretation. The size of the cluster depends on many body excitations controlling the mean free path of the exited photoelectron inside the material and the core hole lifetime. The lifetime of the final state is in the femtosecond range therefore the final states probe the structure of the nano-cluster.

The La $L_3$-edge XANES is determined by dipole transitions from the La($2p_{3/2}$) initial state (n=2, L=1, J=3/2) of the photoabsorber La atom in the spacer layers La$_2$O$_{2.06}$ to final states consisting of multiple scattering resonances of the emitted photoelectron with L=2 orbital moment.

The La $L_3$-XANES spectra were calculated using the full multiple scattering (MS) theory based on self-consistent-field (SCF) potential with muffin-tin approximation, as implemented in the FEFF9.0 code. The original F$mmm$ La$_2$CuO$_4$ is used to simulate the size of the cluster that is needed to reproduce all experimental XANES spectrum. The SCF radius is 5 Å and the FMS radius is 8 Å, which is sufficient to achieve a good convergence. The Hedin-Lundqvist exchange-correlation potential is used. We have found that the cluster probed by La $L_3$-edge XANES, shown in Fig. 3, is made of 21 moles of La$_2$CuO$_4$ with a radius of 0.8 nm.

Previous µXRD experiments [62,63] evidenced the presence of two different phases coexisting in the sample. The first consists of nanoscale OPP with a stoichiometric oxygen content and the second of nanoscale ORP with an oxygen formal content 4.125.

The calculations for the expected XANES spectrum for OPP have been carried out using a cluster with the symmetry F$mmm$ La$_2$CuO$_4$ and the XANES spectrum for ORP has been calculated using the local structure of the doped La$_2$CuO$_{4.125}$. To account for the ORP with O$i$ located in the 1/4,1/4,1/4 site, and forming compositional stripes in the La$_2$O$_2$ structure, we have to include three O$i$ in the atomic cluster probed by the photoabsorption final state around the La absorbing atom in the spacer La$_2$O$_2$ layers as shown in Fig. 3.

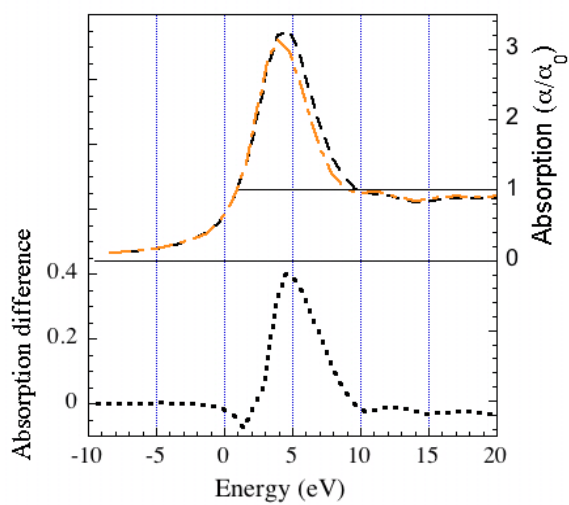

**Figure 4. Upper panel.** XANES calculation at the La $L_3$ using the cluster shown in Figure 3 for the La$_2$CuO$_4$ (brown dashed line) and the La$_2$CuO$_{4.125}$ (black dashed line). **Lower panel.** The absorption difference between the two spectra.

The results of the calculations are shown in Fig. 4. The presence of O$i$ induces an increase of the peak A in Fig. 1 due to the first and strongest MSR as shown in the lower panel of Fig. 4. Figure 4 (upper panel) shows the superimposition of two XANES spectra in the ORPs i.e., the La$_2$CuO$_{4+y}$ with y = 0.125 O$i$, and in the OPPs, i.e., the La$_2$CuO$_4$ without O$i$ (y = 0), computed with the FMS theory. The graph shows the increase of the peak A by 0.4 in units of the La $L_3$-edge atomic absorption jump and a shift of the white-line peak in correspondence of the ORP. The calculation has been performed using the cluster shown in Fig. 3, both for the La $L_3$-edge spectrum of the La$_2$CuO$_{4.125}$ and of the La$_2$CuO$_4$ cluster. The lower panel of Fig. 4 shows the difference between the two spectra. Because the atomic component is removed by this procedure, the presence of the additional O$i$ in proximity of the edge is a mark of the increase of the MS resonance at 5 eV above the threshold.

The XANES spectrum recorded over a micron size spot is already the averaged ratio between ORP and OPP in the same illuminated spot. As a consequence, the underdoped sample (y = 0.06) show a content of O$i$ intermediate between the La$_2$CuO$_4$ and the optimally doped La$_2$CuO$_{4.125}$ samples, which we considered for the FMS calculations.

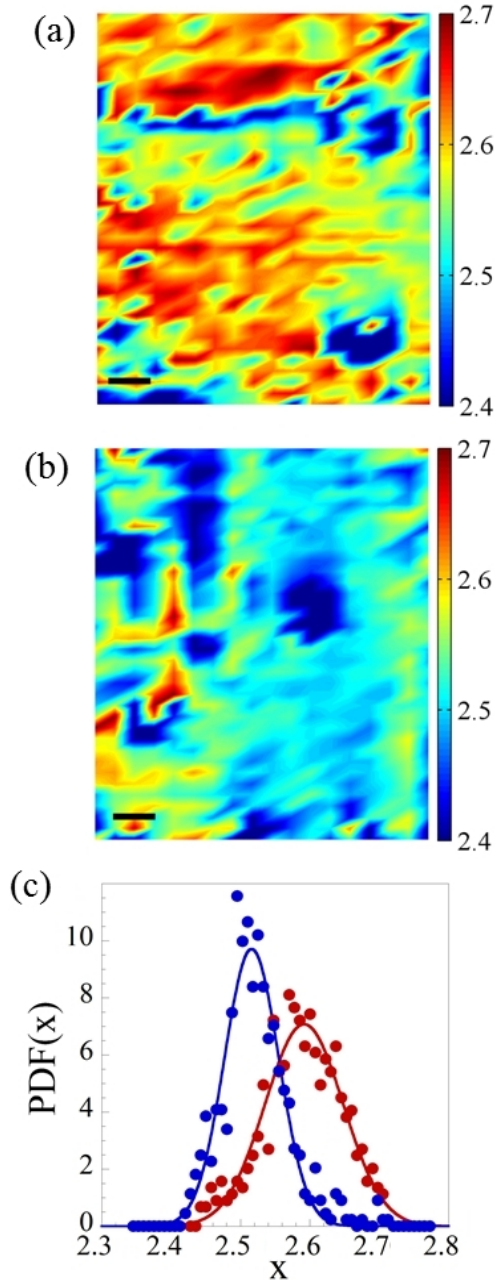

**Figure 5. a, b** Colormaps of the position dependent intensity of the MSR in the XANES spectra, of the investigated La$_2$CuO$_{4.06}$ crystal (intensity is depicted in the colorbar). The maps are 36 x 18 pixels (4 μm pixel size). The black bar corresponds to 10 μm. The exposition time is 10 s/step;**c)** it shows the probability density function PDF(x) of the MSR of the two regions.

Having established the relation between the variation of the main MSR peak of the XANES spectra and the presence of O$i$ in the illuminated spots, we have measured the variation of the intensity of the main MSR peak in the XANES spectra using the X-ray spot at 4 μm steps. The distribution of ORP and OPP is revealed by the map of the space variation of the MSR, peak A, in Fig. 5, where we compare images of two regions with different distributions of ORP.

The statistical analysis of the two regions of the sample is shown in Fig. 5c. The x axis in figure 5c is the intensity of the MSR, which ranges from 2.4 to 2.7 in units of the absorption jump. The distribution measured for the upper panel (5a) is peaked at 2.6, while the lower panel (5b) is at 2.5. Both regions display a Gaussian distribution. The distribution curves of the color-map in panel 5a (5b) are plotted in panel c with red (blue) color dots. The width of the oxygen poor distribution ranges from 2.45 to 2.55, while in the oxygen rich region ranges from 2.52 to 2.65, showing a different distribution of ORP in the crystal and in different locations. The variation of the MSR intensity is ~0.3 instead of the calculated value of 0.4 since using a micron-size beam it is impossible to probe a region free from ORP. This claiming is further supported by scanning μXRD experiments in a similar crystal [63], showing that region without ORP are < 4 μm.

The results confirm the presence of an intrinsic phase separation in a single crystal of an underdoped La$_2$CuO$_{4+y}$, revealing in addition the coexistence of ORP and OPP, characterized by a stronger and a weaker MSR peak, respectively.

The maps in Fig. 5 show clearly the scenario of percolative superconductivity [30-54,87,88]. In order to understand the effect of the percolative scenario on the superconductivity, we have measured the complex surface resistivity of the sample (see Fig. 6).

The surface complex resistivity experiment was performed to measure the superconducting critical temperature. The single-coil inductance technique, non-destructive and contactless, enabled to measure the complex conductivity of the same crystal surface investigated by X-Ray diffraction. A spiral coil of 0.5 mm in diameter with an inductance L, was placed on the sample surface illuminated by x-rays. A change of impedance of the LC circuit was detected as a change of resonant frequency ω in the range 2–4 MHz, and of the amplitude V of the oscillating signal. The high sensitivity of the method arises from the strong mutual inductance between sample and coil in the single-coil geometry. The temperature variation of the ratio of the resonant frequency and the reference frequency of the coil measured in the proximity of the non-superconducting copper metal is a measure of the surface resistivity vs. frequency, proportional to the London penetration depth.

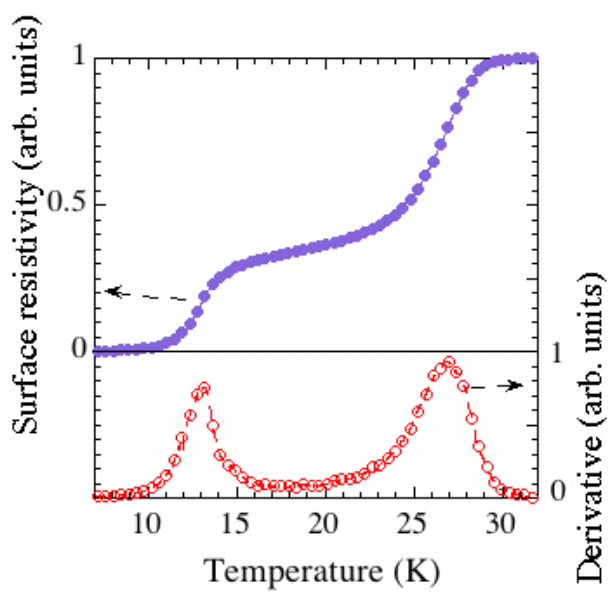

**Figure 6.** The temperature-dependent complex surface resistivity of the underdoped $La_2CuO_{4+y}$ crystal (blue line) and its derivative (red line) showing two critical temperatures $T_{c1}$ = 27 K and $T_{c2}$=13 K. We used a single-coil inductance method recording $(\omega_o/\omega(T))^2$, where $\omega(T)$ is the resonance frequency of an LC circuit, L is the inductance of the coil placed near the surface of the superconducting sample and $\omega_0$ is the reference resonance frequency for a non-superconducting sample.

The surface resistivity is shown in Fig. 6. The results provide the critical temperature ($T_c$) evidence of a first resistivity jump at $T_{c1}$=27 K and a second resistivity jump at $T_{c2}$=13 K, in agreement with previous data [63,89]. The curve illustrates a two-step resistance vs. temperature behavior similar to the case of superconductivity in a network of superconducting islands [90].
In this network, the transition marked as $T_{c1}$ is assigned to intra-island superconductivity, although inter-island phase coherence is still lacking.

Decreasing the temperature, the resistivity also decreases as the coherence length increases up to reach the phase where it becomes comparable to the island spacing. The system evolves toward a global phase coherence, undergoing a transition to a superconducting state at $T_{c2}$ where the entire system becomes superconductor.
Actually, the $T_{c2}$ value depends on the average distance between ORP and is expected to have a finite value for distances among grains of the order of nanometers. Therefore we assign the two jumps in the resistivity plot to the portion of the crystal showed in Fig. 5a. For the portion of the crystal in Fig. 5b, the distance between ORP is too large to give a finite $T_{c2}$.

In conclusion, we have reported scanning μXANES investigation for an underdoped cuprate oxide with mobile O*i* dopants, showing the spatial distribution of ORP on a micron scale. The μXANES technique is perfectly suited to visualize the nature of these multi-scale systems, being a fast and local probe of a selected atomic species. The results presented are complementary to other techniques such as scanning μXRD [62,63], and provide further support to the scenario of percolative superconductivity in copper oxides and other high temperature cuprates superconductors. Moreover, the arrangement of ORP in the real space shows a compelling evidence of different percolation regimes occurring in this system. The percolation phenomenon has a strong influence in the superconducting properties measured, as shown by the occurrence of two critical temperatures and is of support several proposed models that consider the intrinsic percolative nature of superconductivity [30-54,88,89].

**Acknowledgments**
The work was supported by the Dutch FOM and NWO foundations. N.P. acknowledges the Marie Curie Intra European Fellowship for financial support. W.X. acknowledges National Natural Science Foundation of China (Grant No. 11105172). A.B and N.P. acknowledge Superstripes Institute, and European CALIPSO TransNational Access Program of programs of SOLEIL for financial support.